%% file: main_arxiv.tex
\documentclass[10pt,journal]{IEEEtran}

\usepackage{amsmath,amssymb}
\usepackage{graphicx}
\usepackage{url}
\usepackage{array}
\usepackage{booktabs}
\usepackage[table]{xcolor}
\usepackage{enumitem}
\usepackage{tikz}
\usetikzlibrary{shapes.geometric, arrows, positioning, calc}
\usetikzlibrary{decorations.pathreplacing, calligraphy}
\usepackage{caption}
\usepackage{graphicx} 
\usepackage{subcaption} 
\usepackage[acronym]{glossaries}
\glsdisablehyper
\makeglossaries
\input{acronyms}

\usepackage[markup=underlined]{changes} 
\definechangesauthor[name={Ferran}, color=blue]{FB} 
\definechangesauthor[name={Hamid}, color=red]{HL} 

\usepackage[colorlinks,urlcolor=blue,linkcolor=blue,citecolor=blue]{hyperref}

\begin{document}

\title{Cybersecurity in Power Grids: Standards and Research Challenges}
\author{
Ferran~Bohigas-Daranas\IEEEauthorrefmark{1},
Hamid~Latif-Mart\'inez\IEEEauthorrefmark{2},
Nicol\'as~Llorens\IEEEauthorrefmark{3},
David~Bru~i~Bru\IEEEauthorrefmark{4},
Oriol~Gomis-Bellmunt\IEEEauthorrefmark{1},
Eduardo~Prieto-Araujo\IEEEauthorrefmark{1},
Pere~Barlet-Ros\IEEEauthorrefmark{2}\IEEEauthorrefmark{3}%
\thanks{\IEEEauthorrefmark{1}CITCEA, Universitat Polit\`ecnica de Catalunya, Barcelona, Spain.}%
\thanks{\IEEEauthorrefmark{2}BNN, Universitat Polit\`ecnica de Catalunya, Barcelona, Spain.}%
\thanks{\IEEEauthorrefmark{3}Hypergraph, Barcelona, Spain.}%
\thanks{\IEEEauthorrefmark{4}iGrid T\&D, Barcelona, Spain.}%
\thanks{Corresponding author: Ferran Bohigas-Daranas (ferran.bohigas@upc.edu).}%
\thanks{This work was supported by HP2C-DT project (European Union NextGenerationEU/PRTR TED2021-130351B-C21), BLOSSOMS project (grant PID2024-158530OB-I00 by MICIU/AEI/10.13039/501100011033/ and ERDF/EU) and GRAPHS4SEC project (PCI2023-145974-2 by MICIU/AEI/10.13039/501100011033). This work is also partially funded by ICREA. }%
}

\maketitle


\begin{abstract}
This paper examines Smart Grid cybersecurity, emphasizing the critical distinctions between IT and OT environments. It analyzes grid architecture, substation threats, and key international standards, specifically IEC 62351, IEC 62443, and ISO 27001. Finally, it overviews latest research trends, including AI-driven threat detection.
\end{abstract}


\glsresetall

\section{Introduction}
The electrical grid is often described as the most complex machine ever built. It is a critical infrastructure essential for the functioning of modern society. Historically, the grid was an isolated system, relying on proprietary protocols and physical separation for security.  However, the demand for renewable energy integration, real-time monitoring, and efficiency has driven the digitization of the sector, with the extensive use of communication protocols such as IEC 60870-5-104 and DNP3, which rely on TCP/IP networks, and IEC 61850, whose messaging profiles include both TCP/IP-based (MMS) and Layer-2 Ethernet (GOOSE, SV) communications, making the grid a target for a wide range of cyberattacks \cite{tuyen2022comprehensive}. Although many cybersecurity techniques from \gls{it} environments are applicable to power systems, the operational constraints of electrical grids require a fundamentally different prioritisation framework. In \gls{ot} environments, availability and safety often take precedence over confidentiality. Moreover, cybersecurity incidents may translate directly into physical consequences such as equipment damage or large blackouts. As a result, protecting the electrical grid requires a cyber-physical perspective that combines communication security with operational resilience.

This article provides an overview of the cybersecurity challenges affecting modern electrical grids. It addresses the unique challenges of securing industrial control systems, which differ fundamentally from standard corporate \gls{it} networks. We examine the structure of the grid, the motivation for cyberattacks, and the framework required for a robust defense, with a specific focus on the technical standards \gls{iec} 62351 and \gls{iec} 62443.

\section{The Electrical Grid Architecture}
To understand the attack surface, one must first understand the physical and logical structure of the grid. The electrical system is divided into several distinct domains, all of which are increasingly interconnected.

\begin{figure*}[htbp]
     \centering
     \begin{subfigure}[b]{0.65\textwidth}
         \centering
         \includegraphics[width=\textwidth]{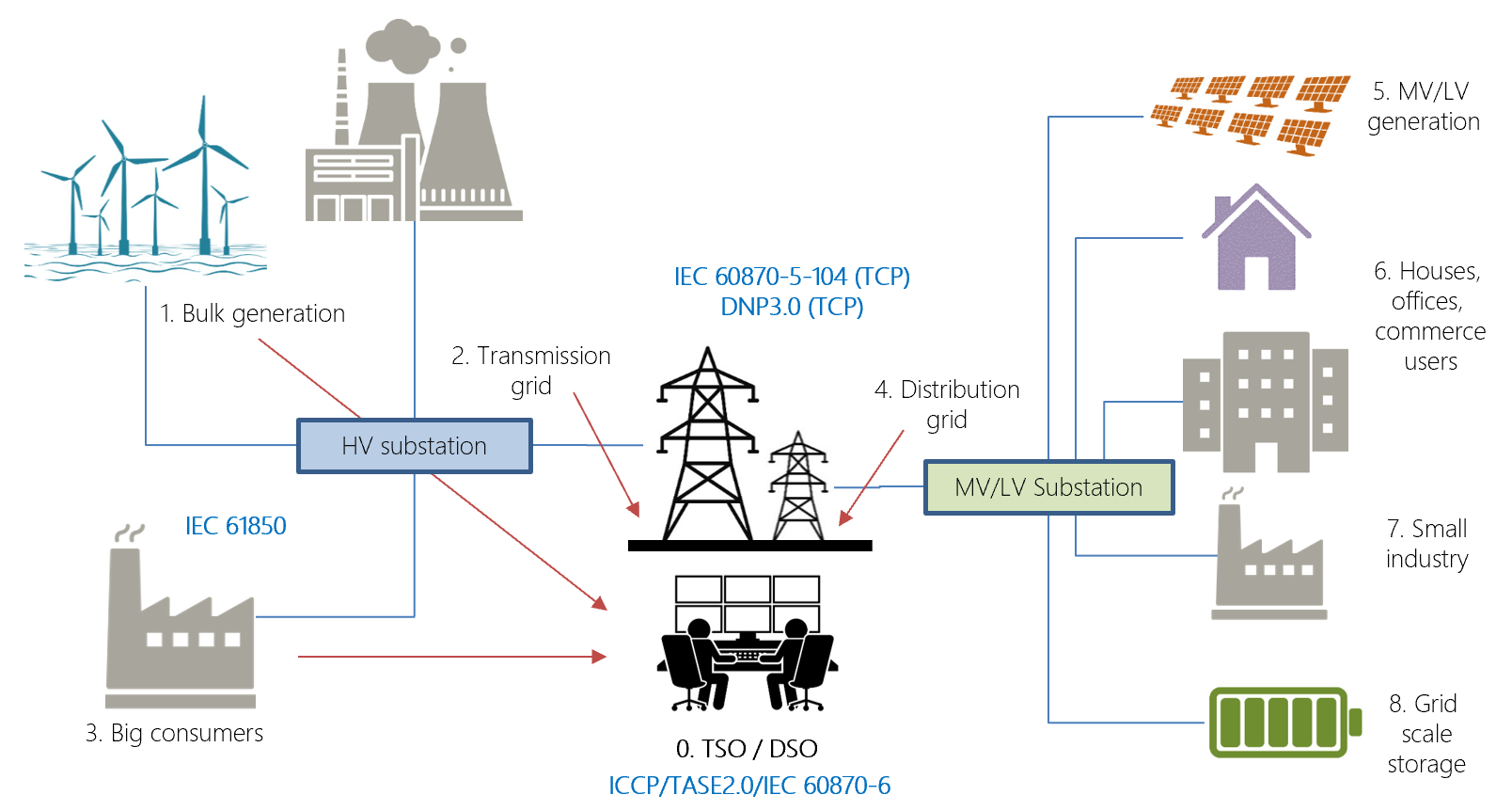}
         \caption{Electrical Grid Scheme}
         \label{fig:ElectricalGridScheme}
     \end{subfigure}
     \hfill 
     \begin{subfigure}[b]{0.3\textwidth}
         \centering
          \resizebox{\textwidth}{!}{                            \input{tikz/purdue.tex} }         
         \caption{The Purdue Reference Model}
         \label{fig:purdue_model}
     \end{subfigure}
     \caption{Electrical grid and Purdue model.}
     \label{fig:total_figure}
\end{figure*}

The electrical grid operates through a hierarchy of functions (Fig.\ref{fig:ElectricalGridScheme}):
\begin{enumerate}
     \item \textbf{Bulk generation:} This includes large-scale power plants such as nuclear, hydro, and thermal facilities that generate high-voltage electricity.
     \item \textbf{Transmission grid:} The high-voltage network responsible for transporting energy over long distances to substations.
     \item \textbf{Distribution grid:} The medium and low-voltage network that delivers electricity from substations to end-users, but which also receives the generation from smaller renewable facilities.
     \item \textbf{Consumption:} This ranges from big industrial consumers to small residential houses, offices, electric vehicle chargers, including the increasing number of roof PV generation.
\end{enumerate}

The operation and coordination of these domains are carried out by multiple entities with clearly defined responsibilities, reflecting the functional separation of the grid itself:
\begin{itemize}
     \item \textbf{\protect\gls{tso}:} Responsible for the stability and operation of the high-voltage transmission network.
     \item \textbf{\protect\gls{dso}:} Responsible for the medium and low-voltage distribution networks.
    \item \textbf{Generation plant owners:} Responsible for the operation and maintenance of generation plants of different nature (nuclear, gas, wind farms, photovoltaic plants, ...)
\end{itemize}

As the grid modernizes, technologies such as distributed generation (solar panels on roofs) and grid-scale storage are blurring the lines between these domains. This increased interconnection expands the potential entry points for cyberattacks and motivates a holistic view of the grid architecture when analyzing its security.

\subsection{Logical Segmentation: The Purdue Model}

To manage the convergence of IT and OT, the industry has historically used the Purdue Reference Model for Industrial Control Systems (Fig.\ref{fig:purdue_model}) to organize assets into a hierarchical structure of levels based on their function and security requirements:
\begin{itemize}
    \item \textbf{Level 4 \& 5 (Enterprise Zone)}: This is a standard IT environment subject to conventional IT security practices, with corporate applications (ERP, billing,...). 
    \item \textbf{Level 3.5 (Industrial DMZ)}: A security buffer zone, not present in the original model but widely adopted in practice, that enforces controlled separation between the enterprise and operational layers, which ensures that no direct connection exists between them.
    \item \textbf{Level 3 (Operations Support)}: This level is responsible for operations scheduling and performance monitoring, and represents a critical boundary between the IT and OT domains.
    \item \textbf{Level 2 (Control Level)}: The SCADA system used by operators to supervise the physical process in real time. Commands issued at this level directly affect grid equipment.
    \item \textbf{Level 1 (Field Control)}: The Intelligent Electronic Devices (IEDs), Programmable Logic Controllers (PLCs), and Remote Terminal Units (RTUs) that execute control logic and relay measurements upward.
    \item \textbf{Level 0 (Physical Process)}: The actual grid equipment, such as transformers, circuit breakers, sensors, and actuators, whose state is monitored and controlled by the layers above.
\end{itemize}
By aligning the grid's architecture with these levels, operators can implement the "Zones and Conduits" segmentation model mandated by IEC 62443. The primary security objective is to ensure that a compromise at the Enterprise Zone (Levels 4–5) cannot propagate laterally into the Control Zones (Levels 0–2), with the Industrial DMZ (Level 3.5) serving as the enforced chokepoint.
The Purdue Model is increasingly regarded as a baseline rather than a definitive architecture for modern deployments. As grid infrastructure becomes more interconnected and cloud-dependent, the strict hierarchical boundary assumptions of the Purdue Model are difficult to maintain in practice. The IEC 62443-3-2 risk-based zone-and-conduit methodology provides a more flexible and asset-centric approach.

A further consideration arising from grid modernisation is the emergence of new operational entities that sit outside the traditional TSO/DSO/generator ownership model. Aggregators and Virtual Power Plant (VPP) operators coordinate large numbers of distributed energy resources, such as small PV plants, battery storage systems and demand-response assets, across both the distribution and consumption domains. These entities communicate with individual DER assets via the Internet and with system operators via energy market interfaces, creating communication paths that span multiple Purdue levels and bypass traditional OT network boundaries entirely. From a cybersecurity standpoint, aggregator platforms represent a significant and growing attack surface that warrants explicit security treatment under IEC 62443 zone-and-conduit analysis.

\section{Cybersecurity requirements and objectives}

In addition to the new attack vectors introduced by grid modernization, the fact that power systems are classified as critical infrastructure further justifies the investment required to guarantee cybersecurity in power grids. A successful cyberattack does not merely result in data loss; it may cause physical damage to equipment, widespread blackouts, economic disruption, and even threats to human safety \cite{presekal2025cyberattacks}.

Historically, \gls{ot}, the hardware and software responsible for monitoring and controlling physical processes, was physically isolated from external networks. However, modern power systems increasingly interconnect \gls{ot} with \gls{it} infrastructures to enable remote maintenance, billing, and large-scale data analysis. This convergence significantly expands the attack surface, exposing legacy \gls{ot} components, often designed without security considerations, to contemporary internet-based threats.

From a cybersecurity perspective, protection mechanisms are commonly structured around the core security objectives of confidentiality, integrity, and availability, collectively known as the CIA triad \cite{gunduz2020cyber}. While these principles apply to both \gls{it} and \gls{ot} systems, their relative prioritization differs substantially between the two environments. This reflects the distinct operational constraints and risk profiles of cyber-physical infrastructures.

In a corporate \gls{it} environment (e.g., banking or email), the priorities are usually:
\begin{enumerate}
    \item \textbf{Confidentiality:} Protecting data privacy is the biggest concern.
    \item \textbf{Integrity:} Ensuring data is not altered.
    \item \textbf{Availability:} Ensuring systems are up and running.
\end{enumerate}

In a real-time OT environment such as an electrical substation, however, these priorities are inverted:
\begin{enumerate}
    \item \textbf{Availability:} The power must stay on. Real-time communication (often in milliseconds) is critical.  A delay of seconds can cause equipment damage or grid instability.
    \item \textbf{Integrity:} To mitigate the risk of physical damage or erroneous decision-making, it is essential to ensure the integrity of both control commands and field measurements.
    \item \textbf{Confidentiality:} While is less immediately safety-critical than availability or integrity, regulatory frameworks such as NERC CIP impose significant requirements on the protection of sensitive operational data.
\end{enumerate}

\begin{table*}[ht]
\centering
\caption{Mapping of CIA Triad to Grid Attacks and Mitigating Standards}
\label{tab:cia-mapping}
\renewcommand{\arraystretch}{1.4} 
\small 

\begin{tabular}{>{\centering\arraybackslash\bfseries}m{2.8cm} >{\centering\arraybackslash}m{4.2cm} m{6.5cm}}
\toprule
\rowcolor{gray!15} 
\gls{cia} Pillar & Typical Attacks & Mitigating Standards \& Controls \\ 
\midrule

Confidentiality & Eavesdropping \par Traffic Analysis \par Sniffing & 
\begin{itemize}[leftmargin=*, nosep, after=\vspace{-\baselineskip}, before=\vspace{-0.5\baselineskip}]
    \item \textbf{IEC 62351-3}: TLS for TCP/IP profiles.
    \item \textbf{ISO 27001}: Data classification policies.
    \item \textbf{IEC 62443-4-2}: Device-level protection.
\end{itemize} \\ \addlinespace

\midrule

Integrity & Man-in-the-Middle (MitM) \par False Data Injection (FDI) \par Replay attacks & 
\begin{itemize}[leftmargin=*, nosep, after=\vspace{-\baselineskip}, before=\vspace{-0.5\baselineskip}]
    \item \textbf{IEC 62351-4/5/6}: Digital signatures.
    \item \textbf{IEC 62443-3-3}: System authenticity.
\end{itemize} \\ \addlinespace

\midrule

Availability & DoS/DDoS \par Ransomware \par Protocol flooding & 
\begin{itemize}[leftmargin=*, nosep, after=\vspace{-\baselineskip}, before=\vspace{-0.5\baselineskip}]
    \item \textbf{IEC 62443-2-1}: Incident response.
    \item \textbf{ISO 27001}: Business Continuity (BCP).
    \item \textbf{IEC 62351-7}: Security event monitoring and network management.\par
\end{itemize} \\ 

\bottomrule
\end{tabular}
\end{table*}

In addition to the CIA triad, there are two Extended Security Goals for power grid cybersecurity. \textbf{Accountability} refers to the ability to determine, for every action, who performed it, from where, when, and why. This ensures that forensic analysis of any event is supported by maximum information, aligned with ISO 27001 standards and the \gls{pdca} cycle described later in this paper. A second extended goal is \textbf{Resilience}, just as important as protection is the ability to recover the entire system following an attack. A robust backup policy is required to allow system operators to restore applications, configurations, and credentials, ensuring a timely return to normal operations.

\section{Threats and Vulnerabilities}
The grid faces threats from a range of adversaries, including nation-states, hacktivists, and criminal organizations, who exploit vulnerabilities specific to industrial control environments, which need to be addressed by applying known techniques, some of them already included in the cybersecurity-related standards, such as \gls{iec} 62351, \gls{iec} 62443, and ISO 27001.

\subsection{Electrical Grid Cyber Vulnerabilities}
The communication network, which is supervising and controlling the electrical grid, has some unique characteristics that can be seen as vulnerabilities:
\begin{itemize}
    \item Vast extension, with sites located in remote and isolated locations.
    \item Unmanned sites, which work autonomously.
    \item A heterogeneous mix of legacy and modern devices; older equipment often cannot be updated remotely and lacks the computational resources to support encryption or authentication.
    \item Reliance on legacy protocols that lack built-in cybersecurity mechanisms.
    \item Critical infrastructure working in real time, usually without the option to stop them for long periods.
    \item Authentication management and certificate lifecycle in OT environments present unique operational challenges. Unlike IT networks, where certificate authorities and identity management systems can be centrally administered and frequently updated, OT environments often comprise isolated or semi-air-gapped network segments with constrained connectivity. Renewing certificates, or revoking compromised identities in these settings may require physical access to remote sites, creating windows of exposure. Furthermore, many legacy IEDs lack support for standard \gls{pki} mechanisms, making consistent identity verification an unsolved operational problem.
\end{itemize}
\subsection{Types of Attacks} \label{lab:Types-of-attacks}
Cyber attacks are likely to disrupt the CIA triad by compromising data confidentiality, undermining the integrity of grid information, or preventing the availability of resources.
Common attack vectors in the electrical sector include:
\begin{itemize}
    \item \textbf{Man-in-the-Middle (Confidentiality):} An attacker intercepts communication between the control center (SCADA) and the substation, potentially altering commands, by breaking the confidentiality.
    \item \textbf{Credential Theft (Confidentiality):} Obtaining credentials to access the \gls{ot} network, or exploiting software update mechanisms to introduce backdoors.
    \item \textbf{Spoofing (Integrity/Confidentiality):} An attacker impersonates a legitimate device or control center by forging its network identity, protocol address, or cryptographic credentials. In \gls{ot} environments, where many legacy protocols such as IEC~60870-5-104 or DNP3 lack built-in sender authentication, a spoofed message is indistinguishable from a genuine command. This allows an adversary to issue unauthorized switching operations or inject falsified measurements while appearing to originate from a trusted source, bypassing perimeter defenses that rely solely on network address verification.
    \item \textbf{False Data Injection or FDI (Integrity):} Inserting false data, such as wrong measurements or false topological information, into the \gls{ot} network to confuse the decisions taken by Control Center operators.
    \item \textbf{Denial of Service/Distributed Denial of Service or DDoS (Availability):} Flooding the network with traffic to prevent legitimate control commands from reaching field devices, or to block measurements, alarms, and telemetry from remote stations from reaching the control center.
    \item \textbf{Malware Injection:} Introducing malicious software (like ransomware) into the control network via USB drives or phishing emails.
    \item \textbf{Lateral Movement and Privilege Escalation}: Once initial access is gained, attackers navigate from lower-security zones toward critical control zones. In \gls{ot} environments, compromising administrative credentials allows attackers to modify \gls{ied} configurations or disable protections while remaining undetected by perimeter defenses \cite{Paul2024}.
\end{itemize}

These attacks can be deployed broadly against any corporation to extract ransom payments; however, they can also be specifically orchestrated by nation-states or organizations targeting energy infrastructure to undermine a country's stability through cyberwarfare or terrorism. This threat has prompted various nations to establish dedicated security organizations and frameworks to protect their critical infrastructure from cyber risks. Examples include ENISA in the European Union, supported by the \gls{cra}, and \gls{nist} in the United States.

\subsection{Case Studies}
Despite the existence of operational and regulatory safeguards, several incidents have demonstrated that cyberattacks threats to power systems are not merely theoretical, but a reality that grid operators must face in their daily operation. 

\subsubsection{Industroyer (CrashOverride)}
The Industroyer malware \cite{cherepanov2017win32}, also designated CrashOverride, represents the most technically sophisticated cyberattack on electrical infrastructure documented to date. It followed the 2015 BlackEnergy 3/Sandworm \cite{njccic_blackenergy_nodate} campaign, which manually compromised IT networks to cut power to 230,000 customers in Ukraine. The 2016 Industroyer attack marked a qualitative escalation, engineered to operate autonomously at the control level. On 17 December 2016, it triggered a power outage in Kyiv, affecting approximately 225,000 customers for one hour.

Unlike BlackEnergy, Industroyer contained dedicated protocol-aware payloads capable of interacting directly with substation automation equipment. It utilized standard industrial telecontrol protocols, including IEC 60870-5-104, IEC 60870-5-101, IEC 61850, and OPC DA, to issue commands directly to switchgear. This modularity suggests the malware was designed for reuse across diverse utility environments. The attack was structured across four distinct components:
\begin{itemize}
    \item \textbf{Protocol-level switchgear control}: The primary payload transmitted valid IEC 60870-5-104 command frames to open circuit breakers. Because these commands were syntactically legitimate, they were indistinguishable from authorized actions by protocol-unaware monitoring tools.
    \item \textbf{Denial of service against protection relays}: This component targeted relays responsible for fault isolation. By rendering them unresponsive, attackers sought to prevent automatic restoration and complicate manual re-energisation.
    \item \textbf{Configuration destruction}: Attackers corrupted IED configuration files and relay settings, forcing operators to reconstruct device data before the substation could safely return to service, prolonging recovery time.
    \item \textbf{Wiper payload (KillDisk)}: The KillDisk module overwrote Master Boot Records on engineering workstations. By eliminating human-machine interfaces, it prevented operators from assessing the substation state and forced a physical recovery process.
\end{itemize}
Industroyer illustrates a coherent strategy: cause an outage, then systematically disable the tools needed to restore it, evidencing deep knowledge of grid restoration procedures.

\subsubsection{Dragonfly}
The Dragonfly campaign (or Energetic Bear), active from 2011 to 2014, initially targeted energy companies for strategic intelligence. Using spear-phishing and watering-hole attacks, actors compromised IT systems to harvest SCADA configurations without seeking immediate disruption \cite{wueest2014dragonfly}. 

A second phase, Dragonfly 2.0 (2015–2017), achieved access to operational systems in the US and Europe. US-CERT Alert TA18-074A \cite{cisa_russian_2018} confirmed attackers reached the Stage 2 ICS environment, obtaining the credentials necessary for disruptive operations. As no disruption was triggered, Dragonfly 2.0 is characterized as a pre-positioning campaign for future conflict. Together, these phases illustrate a pattern of patient reconnaissance followed by deep intrusion with disruption held in reserve.

\subsubsection{sPower: First Grid-Disruptive Cyber Event}
On March 5, 2019, sPower became the target of a cyberattack exploiting vulnerabilities in web-facing firewalls \cite{sobczak_first---kind_2019}. This Denial-of-Service (DoS) attack triggered continuous reboots of security appliances, resulting in communication outages lasting over 12 hours. While the attack did not cause physical load shedding, it disrupted command-and-control links across Utah, Wyoming, and California, creating a 500 MW "blind spot" in grid visibility. These incidents have motivated the development of the dedicated security standards described in the following section.

\section{Security Standards for Power Systems}

The increasing exposure of power systems to cyber threats has driven the development of dedicated industrial standards and solutions aimed at improving the security of operational environments. In line with this, the energy sector has progressively adopted international standards and regulatory frameworks to secure communications, define security requirements, and ensure interoperability across vendors. One of the main focuses has been communication security, as many cyberattacks exploit weaknesses at the protocol and network levels.

\subsection{Strategies and Techniques for Attack Mitigation}

The cyber-physical threats detailed in Section \ref{lab:Types-of-attacks} can be mitigated or neutralized by implementing a defense-in-depth strategy. The following techniques are essential for securing grid infrastructure:

\begin{itemize}
    \item \textbf{Data Encryption}: This involves transforming plaintext into ciphertext via cryptographic algorithms. In a grid context, encryption ensures \textit{confidentiality} and \textit{integrity} for sensitive telemetry data. Furthermore, it facilitates \textit{non-repudiation} through digital signatures and certificates, ensuring that commands sent to \gls{ied} originate from authenticated control centers.
    
    \item \textbf{\protect\gls{rbac}}: \gls{rbac} restricts system access based on the specific responsibilities of users within the utility's organizational structure. By adhering to the "principle of least privilege", it prevents unauthorized personnel or compromised accounts from executing critical switching operations or accessing sensitive configuration files.

    \item \textbf{Firewalls and Network Segmentation}: These systems monitor and filter traffic based on predefined security policies. In electrical architectures, firewalls serve as vital logical barriers between the corporate IT network and the high-availability OT network (the "Electronic Security Perimeter"), preventing lateral movement by attackers.

    \item \textbf{Continuous Staff Training}: The human element remains a primary vulnerability. Training programs mitigate risks such as social engineering and phishing. For grid operators, this also includes protocol-specific training to ensure security policies are maintained during emergency manual overrides or maintenance windows.

    \item \textbf{Intrusion Detection Systems (IDS)}: Given the deterministic nature of power system communications (e.g., Modbus, DNP3, or \gls{iec} 61850), an \gls{ids} can monitor network traffic for anomalies or known attack signatures. This provides real-time visibility into the network, allowing for rapid response before an intrusion escalates into a physical outage.
\end{itemize}

These technical measures are codified within international standards, such as \gls{iec} 62351, \gls{iec} 62443, and ISO/\gls{iec} 27001, which are discussed in the following section

  

\subsection{IEC 62351: Securing Communications}
One of the primary pillars of grid security is the \gls{iec} 62351 standard family \cite{iec62351}, developed by \gls{iec} Technical Committee 57 (TC57). Its main objective is to provide security mechanisms for the specific protocols used in the energy sector, including \gls{iec} 60870-5 (Telecontrol), \gls{iec} 60870-6 (ICCP for inter-control center communication), \gls{iec} 61850 (Substation Automation), IEEE 1815 (DNP3) and  CIM (\gls{iec} 61970 \& 61968).

While other standards address general management or internal system security, \gls{iec} 62351 specifically targets the security of the communication protocols used in power system automation. The standard aims to ensure four key security properties for data transfer:
\begin{enumerate}
    \item \textbf{Confidentiality:} Data is encrypted to prevent unauthorized viewing.
    \item \textbf{Integrity:} The system guarantees that data has not been altered during transit.
    \item \textbf{Availability:} Measures are taken to minimize the risk of Denial-of-Service (DoS) attacks.
    \item \textbf{Authentication:} The system ensures that communicating devices are exactly who they claim to be.
\end{enumerate}

\gls{iec} 62351 applies security controls across different layers of the communication stack to create a secure protocol architecture. At the lower layers (network and transport), the standard defines how to wrap specific protocols (like IEC 60870-5-104) within TLS 1.2 or higher (with TLS 1.3 support added in the 2022 revision) to secure the end-to-end communication, protecting against eavesdropping and simple injection attacks. For network access, it integrates with Ethernet security features like MAC filtering and VLANs. At the application layer, \gls{iec} 62351 introduces specific authentication and encryption measures for each protocol, like authentication mechanisms for command execution or encryption to configuration files.

To prevent unauthorized access, \gls{iec} 62351 promotes the use of \gls{rbac}. This limits access rights based on the specific role of the user or device, rather than using generic passwords shared across all staff, and supports centralized authentication management. To achieve this, IEC 62351 relies on \gls{pki}, using Certificate Authorities (CAs) to manage the lifecycle of digital certificates. This ensures that when a device receives a command, it can cryptographically verify that it came from a legitimate control center.

\subsection{IEC 62443: Industrial Automation Security}
While \gls{iec} 62351 focuses on how data moves, \gls{iec} 62443 \cite{iec62443} focuses on the system itself. It is a comprehensive series of international standards for implementing and managing cybersecurity in \gls{iacs} and adopts a "Defense-in-Depth" strategy, meaning multiple layers of security controls are implemented throughout the system to provide redundancy. If one defense (e.g., a firewall) fails, others (e.g., device authentication) are in place to stop the attack.

\gls{iec} 62443 defines seven \glspl{fr} that every secure system must address:
\begin{enumerate}
    \item \textbf{\protect\gls{iac}:} Rigorous management of user and device identities.
    \item \textbf{\protect\gls{uc}:} Enforcing permissions and authorizations so users can only perform actions necessary for their job.
    \item \textbf{\protect\gls{si}:} Protecting against unauthorized modifications to software or firmware.
    \item \textbf{\protect\gls{dc}:} Protecting data from unauthorized disclosure.
    \item \textbf{\protect\gls{rdf}:} Segmenting networks using zones and conduits to prevent unrestricted movement.
    \item \textbf{\protect\gls{tre}:} Detecting, logging, and responding to security events in real-time.
    \item \textbf{\protect\gls{ra}:} Ensuring the system remains operational during stress or denial-of-service attacks.
\end{enumerate}

Furthermore, it applies a "Security by Design" approach, where security is not an add-on but is considered from the initial design phase through implementation, operation, maintenance, and eventual decommissioning.

A core concept of \gls{iec} 62443 is network segmentation to limit the spread of an attack, which is achieved through the use of Zones, logical or physical groupings of assets that share similar security requirements and criticality, and conduits, which are logical groupings of communication requirements, which may be physical (cables, fiber) or logical (VPNs, encrypted tunnels), that allows the passage of data between zones. The standard dictates that the strongest security controls must be placed at the boundaries between zones (in the conduits) to restrict and filter communication.  This prevents an attacker who has compromised a low-security zone (like the corporate office) from easily pivoting to a high-security zone (the substation control room).

 The standard defines \glspl{sl} to quantify the resilience of a system from \gls{sl}~1, which considers only casual or coincidental violations resulting from human error or non-malicious actions, to \gls{sl}~4, protecting against intentional violation using sophisticated means with extended resources and high motivation, executed by nation-states (APTs) with attackers capable of developing zero-day exploits (e.g., Industroyer).

\subsection{Security Management Lifecycle (ISO 27001)}
Implementing cybersecurity is not a one-time event; it is a continuous lifecycle, which is represented in ISO 27001 \cite{iso27001} by the \gls{pdca} cycle, which is based on five key pillars:

 
\begin{itemize}
    \item \textbf{Assessment:} Before implementing defenses, the organization must understand its risks. This involves identifying all assets, finding vulnerabilities, and calculating the probable risk and cost of liabilities.
    \item \textbf{Policy:} This involves creating rules and legal requirements for the security domain. Policies define how security is managed and who has access to what data.
    \item \textbf{Deployment:} This stage involves the practical implementation of the policies. It includes purchasing and installing firewalls, \glspl{ids}, and configuring network segmentation.
    \item \textbf{Training:} Human error is often the weakest link. Continuous training is required for all staff regarding phishing attempts and proper credential handling, as threats are continuously evolving.
    \item \textbf{Audit (Monitoring):} Security is not "set and forget." Auditing is responsible for detecting active security attacks, identifying breaches that have already occurred, and assessing the performance of the installed infrastructure.
\end{itemize}

\subsection{Regulatory Frameworks: Mandatory Compliance and the Voluntary–Mandatory Distinction}
The standards discussed in the preceding subsections (IEC 62351, IEC 62443, and ISO/IEC 27001)  are voluntary international standards: utilities and equipment vendors adopt them because they represent best practice and, increasingly, because procurement contracts or national regulations require them. However, a complete picture of the grid cybersecurity landscape requires acknowledging that, in several major jurisdictions, cybersecurity obligations for energy infrastructure are not voluntary but legally enforceable, and non-compliance carries significant regulatory and financial consequences.

In the European Union, the NIS2 Directive (Directive 2022/2555) \cite{eu_directive_2022} expanded the scope and enforcement rigour of its predecessor, the 2016 NIS Directive. Energy is explicitly listed as a critical sector under NIS2, and system operators of essential services are required to implement risk-management measures covering network security, incident handling, supply chain security, cryptography, and access control. NIS2 also introduces direct personal liability for senior management in the event of non-compliance, a provision intended to elevate cybersecurity from a technical function to a board-level governance concern. For organizations operating within the EU, NIS2 compliance obligations sit alongside and interact with the voluntary IEC standards: the directive does not mandate specific technical standards, but regulators in several member states are beginning to reference IEC 62443 and IEC 62351 as acceptable means of demonstrating compliance.

In North America, the most operationally significant mandatory framework is the NERC CIP (North American Electric Reliability Corporation Critical Infrastructure Protection) \cite{nerc_nerc_nodate} standards series, administered by NERC under authority delegated by the Federal Energy Regulatory Commission (FERC). NERC CIP establishes binding requirements for bulk electric system operators across the United States, Canada, and parts of Mexico, covering areas including asset identification and classification (CIP-002), security management controls (CIP-003), personnel and training (CIP-004), electronic security perimeters (CIP-005), physical security (CIP-006), systems security management (CIP-007), incident reporting (CIP-008), and recovery planning (CIP-009). NERC CIP compliance is an audited obligation: utilities that fail to meet its requirements are subject to financial penalties. Many of the technical controls described in IEC 62443 and IEC 62351 align with NERC CIP requirements, but the compliance burden and audit cadence of the mandatory framework impose operational disciplines that voluntary standards alone do not.

Beyond these two dominant frameworks, other jurisdictions have developed their own mandatory or semi-mandatory regimes, including the UK's NIS Regulations, Australia's Security of Critical Infrastructure Act, and sector-specific requirements in Singapore and Japan, reflecting a global trend towards treating energy cybersecurity as a matter of national legislative interest rather than industry self-regulation.

\section{Emerging Research Directions}

Although \gls{iec} 62351 and \gls{iec} 62443 provide the foundation for securing operational environments, many modern threats require more adaptive and intelligent defensive mechanisms. As a result, recent research has increasingly focused on combining cybersecurity with data-driven and cyber-physical approaches capable of detecting, anticipating, and mitigating attacks in real time. Representative directions include ${i}$ \gls{ml}-based solutions, $(ii)$ \gls{mtd}, and $(iii)$ \gls{cpse}.

\subsubsection{\gls{ml}-based solutions}
Typically, \glspl{ids} rely on heuristic rules, while struggle to adapt to evolving attack strategies or previously unseen traffic patterns. Recent studies address this gap by applying \gls{ml} to model the complex, non-linear relationships present in power grid physics. By learning the normal operational behavior of the grid, these models can flag events that deviate from the expected physical laws, even if this behavior does not violate the predefined heuristic laws.

Early deep-learning approaches demonstrated that anomaly-detection systems could learn temporal operational patterns directly from grid measurements \cite{he2017real}. These methods leveraged architectures such as \gls{cdbn} to extract high-level temporal features from historical operational data, enabling the detection of anomalous behaviors even in the presence of faults or changing operating conditions. More recently, graph-based approaches such as \glspl{gnn} have attracted attention because they can explicitly model the topological structure of the power grid \cite{boyaci2021graph}. By incorporating the spatial dependencies between substations and transmission lines into the learning process, these models improve the detection of attacks such as false data injection, which may remain hidden to approaches that only analyze network traffic or local measurements.

\subsubsection{\gls{mtd}-based solutions}
Contrary to some industry solutions that rely on perimeter-hardening approaches that focus solely on securing access points to the grid, \gls{mtd}-based solutions aim to increase the cost and uncertainty to attackers by continuously modifying selected system parameters or measurements during normal operation. The main objective of these approaches is to invalidate the required knowledge of the system state, transforming cyber-physical attacks that would otherwise remain stealthily into detectable anomalies.

\subsubsection{Cyber-Physical state estimation methods}
\gls{cpse} refers to state estimation techniques that validate the physical plausibility of the grid based on the measurements received from sensors in real time. The main idea behind these techniques is to run a sanity check of the received data: rather than trusting the data solely on the basis of the security of its collection and transmission mechanisms, \gls{cpse} verifies it against the laws of physics, but for this requires a valid physical model. 

Recent \gls{cpse} approaches have extended this idea by treating cyberattacks and anomalous events as disturbances affecting the physical evolution of the grid \cite{su2022cyber}. These methods model the power system as a dynamic cyber-physical process and continuously compare measured observations against predicted system behavior. By enforcing consistency between physical measurements and the expected system dynamics, \gls{cpse} techniques can identify malicious manipulations that preserve measurement-level consistency but violate the physical constraints of the grid. As a result, they provide an additional layer of defense against sophisticated attacks designed to evade conventional detection mechanisms.

\section{Conclusion}
The digitization of the electrical grid is irreversible, offering immense benefits for energy transition and efficiency. However, it binds the physical security of the grid to the digital security of its control systems.

 Several key conclusions must be drawn for any organization operating in this sector:
\begin{enumerate}
    \item \textbf{No absolute security:} No communication-connected system is 100\% secure and there will always be residual risks, the goal is risk management, not only risk elimination.
    \item \textbf{Holistic approach:} Security must be addressed at all levels of the architecture (from physical devices to the enterprise cloud interface) and all levels of the organization (from the engineer to the CEO).
    \item \textbf{Dynamic process:} Cybersecurity is a race between corporate security policies and hostile entities. As attackers evolve, defenses must evolve.
\end{enumerate}

Effective cybersecurity in the power grid requires a shift in mindset, prioritizing availability and safety while implementing rigorous defense-in-depth strategies through standards like \gls{iec} 62351 and \gls{iec} 62443.

\bibliographystyle{IEEEtran}  
\bibliography{references}

\section{Biographies}

\textbf{Ferran Bohigas-Daranas} is a Ph.D. candidate at CITCEA, Universitat Politècnica de Catalunya (UPC), Barcelona, Spain. His research focuses on the use of Graph Neural Networks in electrical engineering, communications, and cybersecurity for the electrical grid. He received an M.Sc. in Electronics Engineering from UPC. Contact him at ferran.bohigas@upc.edu.

\textbf{Hamid Latif-Martínez} is a postdoctoral researcher at the Barcelona Neural Networking Center, Universitat Politècnica de Catalunya (UPC), Barcelona, Spain. His research interests include Graph Neural Networks, network measurement and traffic analysis, and cybersecurity. He received a Ph.D. in Computer Architecture from UPC.

\textbf{Nicolás Llorens} is CEO of Hypergraph AI, Barcelona, Spain. He leads research on graph foundational models for network threat detection and critical infrastructure protection. He received a B.Sc. in Computer Science from Universitat Politècnica de Catalunya (UPC). Contact him at nllorens@hypergraph.tech.

\textbf{David Bru i Bru} is CTO of iGrid T\&D, Barcelona, Spain. His work focuses on real-time systems, communications, and cybersecurity in the electrical system. He received an M.Sc. in Mechatronics from Napier University, UK. Contact him at david.bru@igrid-td.com.

\textbf{Oriol Gomis-Bellmunt} (Fellow, IEEE) is a professor at Universitat Politècnica de Catalunya (UPC), Barcelona, Spain. His research interests include power electronics, power systems, and renewable energy integration in power systems. He received a Ph.D. in Electrical Engineering from UPC. Contact him at oriol.gomis@upc.edu.

\textbf{Eduardo Prieto-Araujo} (Senior Member, IEEE) is an associate professor at Universitat Politècnica de Catalunya (UPC), Barcelona, Spain. His research interests include renewable generation systems, control of power converters for HVdc applications, interaction analysis between converters, and power electronics-dominated power systems. He received a Ph.D. in Electrical Engineering from UPC. Contact him at eduardo.prieto-araujo@upc.edu.

\textbf{Pere Barlet-Ros} is a full professor at Universitat Politècnica de Catalunya (UPC), Barcelona, Spain, and scientific director of the Barcelona Neural Networking Center. His research has been integrated into several open-source and commercial products. He received a Ph.D. in Computer Science from UPC. Contact him at pere.barlet@upc.edu.

\end{document}

%% file: acronyms.tex
\newacronym{iec}{IEC}{International Electrotechnical Commission}
\newacronym{dnp}{DNP}{Distributed Network Protocol}
\newacronym{tcpip}{TCP/IP}{Transmission Control Protocol / Internet Protocol}
\newacronym{it}{IT}{Information Technology}
\newacronym{ot}{OT}{Operational Technology}
\newacronym{tso}{TSO}{Transmission System Operator}
\newacronym{dso}{DSO}{Distribution System Operator}
\newacronym{cia}{CIA}{Confidentiality, Integrity, and Availability}
\newacronym{pdca}{PDCA}{Plan-Do-Check-Act}
\newacronym{ied}{IED}{Intelligent Electronic Device}
\newacronym{enisa}{ENISA}{European Union Agency for Cybersecurity}
\newacronym{cra}{CRA}{Cyber Resilience Act}
\newacronym{tls}{TLS}{Transport Layer Security}
\newacronym{vlan}{VLAN}{Virtual Local Area Network}
\newacronym{goose}{GOOSE}{Generic Object Oriented Substation Event}
\newacronym{sv}{SV}{Sampled Values}
\newacronym{pki}{PKI}{Public Key Infrastructure}
\newacronym{ca}{CA}{Certificate Authority}
\newacronym{sl}{SL}{Security Level}
\newacronym{fr}{FR}{Foundational Requirement}
\newacronym{iac}{IAC}{Identification and Authentication Control}
\newacronym{uc}{UC}{Use Control}
\newacronym{si}{SI}{System Integrity}
\newacronym{dc}{DC}{Data Confidentiality}
\newacronym{rdf}{RDF}{Restricted Data Flow}
\newacronym{tre}{TRE}{Timely Response to Events}
\newacronym{ra}{RA}{Resource Availability}
\newacronym{ids}{IDS}{Intrusion Detection System}
\newacronym{ml}{ML}{Machine Learning}
\newacronym{ai}{AI}{Artificial Intelligence}
\newacronym{mtd}{MTD}{Moving Target Defense}
\newacronym{cpse}{CPSE}{Cyber-Physical State Estimation}
\newacronym{cdbn}{CDBN}{Conditional Deep Belief Network}
\newacronym{gnn}{GNN}{Graph Neural Network}
\newacronym{iacs}{IACS}{Industrial Automation and Control Systems}
\newacronym{rbac}{RBAC}{Role-Based Access Control}
\newacronym{dos}{DoS}{Denial-of-Service}
\newacronym{nist}{NIST}{National Institute of Standards and Technology}
\newacronym{der}{DER}{Distributed Energy Resources}
\newacronym{wan}{WAN}{Wide Area Network}
\newacronym{mitm}{MitM}{Man-in-the-Middle}
\newacronym{fdi}{FDI}{False Data Injection}

%% file: tikz/purdue.tex
\definecolor{itblue}{RGB}{210, 230, 250}
\definecolor{otgreen}{RGB}{210, 245, 210}
\definecolor{dmzorange}{RGB}{255, 235, 200}

\begin{tikzpicture}[
    node distance=1.2cm,
    level_box/.style={rectangle, rounded corners, minimum width=8cm, minimum height=0.8cm, text centered, draw=black, fill=white, font=\small},
    zone_label/.style={font=\bfseries\footnotesize, rotate=90, anchor=center}
]

\node (L5) [level_box, fill=itblue] {Level 5: Enterprise Network (Cloud, External)};
\node (L4) [level_box, fill=itblue, below of=L5] {Level 4: Business Logistics (ERP, Billing, Email)};
\node (L35) [level_box, fill=dmzorange, below of=L4] {Level 3.5: Industrial DMZ (Security Buffer)};
\node (L3) [level_box, fill=otgreen, below of=L35] {Level 3: Operations Systems (MES, Historian)};
\node (L2) [level_box, fill=otgreen, below of=L3] {Level 2: Control Level (SCADA, HMI)};
\node (L1) [level_box, fill=otgreen, below of=L2] {Level 1: Field Control (PLCs, IEDs, RTUs)};
\node (L0) [level_box, fill=otgreen, below of=L1] {Level 0: Physical Process (Sensors, Actuators)};

\draw [thick, decoration={brace, mirror, raise=5pt}, decorate] 
    (L5.north west) -- (L4.south west) node [zone_label, pos=0.5, xshift=+0 cm, yshift=0.6cm] {Enterprise (IT)};

\draw [thick, decoration={brace, mirror, raise=5pt}, decorate] 
    (L3.north west) -- (L0.south west) node [zone_label, pos=0.2, xshift=-1.2cm, yshift=0.6cm] {Industrial (OT)};

\draw [thick, <->] (L5) -- (L4);
\draw [thick, <->] (L4) -- (L35);
\draw [thick, <->] (L35) -- (L3);
\draw [thick, <->] (L3) -- (L2);
\draw [thick, <->] (L2) -- (L1);
\draw [thick, <->] (L1) -- (L0);

\end{tikzpicture}